\documentclass[conference]{IEEEtran}

\usepackage{cite}

\ifCLASSINFOpdf
   \usepackage[pdftex]{graphicx}
\else
   \usepackage[dvips]{graphicx}
\fi

\usepackage{amsmath}
\usepackage[cmintegrals]{newtxmath}

\usepackage{subfigure}

\usepackage{bm}
\usepackage{comment}

\renewcommand{\baselinestretch}{0.85}
\newcommand{\argmax}{\mathop{\rm arg~max}\limits}

\IEEEoverridecommandlockouts
\begin{document}
\title{Layered Belief Propagation for Low-complexity Large MIMO Detection Based on Statistical Beams}

\author{\IEEEauthorblockN{Takumi Takahashi$\,^*$, Antti T{\"o}lli$\,^\dagger$, Shinsuke Ibi$\,^*$, and Seiichi Sampei$\,^*$ }
\IEEEauthorblockA{$\,^*$Department of Information and Communications Technology, Osaka University,\\ Yamada-oka 2-1, Suita 565-0871, Japan\\ 
Email: takahashi@wcs.comm.eng.osaka-u.ac.jp, \{ibi, sampei\}@comm.eng.osaka-u.ac.jp\\
$\,^\dagger$ Centre for Wireless Communications (CWC), University of Oulu, Finland - FI-90014\\
    Email: antti.tolli@oulu.fi
}}

\maketitle
\begin{abstract}
%Root ../main.tex
%===============
% Abstract.
%
This paper proposes a novel layered belief propagation (BP) detector with a concatenated structure of two different BP layers for low-complexity large multi-user multi-input multi-output (MU-MIMO) detection based on statistical beams.
To reduce the computational burden and the circuit scale on the base station (BS) side, the two-stage signal processing consisting of slow varying outer beamformer (OBF) and group-specific MU detection (MUD) for fast channel variations is effective.
However, the dimensionality reduction of the equivalent channel based on the OBF results in significant performance degradation in subsequent spatial filtering detection.
To compensate for the drawback, the proposed layered BP detector, which is designed for improving the detection capability by suppressing the intra- and inter-group interference in stages, is introduced as the post-stage processing of the OBF.
Finally, we demonstrate the validity of our proposed method in terms of the bit error rate (BER) performance and the computational complexity.
\end{abstract}


% \fi
\IEEEpeerreviewmaketitle

\section{Introduction}
%Root ../main.tex

%===============
% Section 1.
%===============
Large multi-user multi-input multi-output (MU-MIMO) systems, in which a large number of antennas are equipped on both the transmitter and receiver sides, have been regarded as one of the most promising technologies for the upcoming fifth generation (5G) cellular systems\cite{Hanzo2009,yang2015,Rusek2013}.
The larger dimensions of MIMO channels can increase data rate with the aid of spatial multiplexing gain and improve detection reliability thanks to the spatial diversity gain, as well as, support a massive amount of simultaneous wireless communications.
%
%Over the past few years, the application of large MIMO techniques is mainly to increase the capacity of the downlink, and many researchers have shown considerable interest in the beamformer design based on the optimization theory.
%
%On the other hand, with the popularization of cloud computing services, uplink large MU-MIMO systems are gaining attention as the data hub for supporting the massive amount of wireless connections \cite{Bockelmann2016}.
%
%However, large MU-MIMO systems make large-scale multi-user detection (MUD) difficult in uplink scenarios, which is computationally expensive \cite{Chockalingam2014}.
%\cite{Datta2013, Li2010, Mohammed2009,cirkic2014}.
%
However, precoder design in the downlink or MU detection (MUD) in the uplink for large MU-MIMO become difficult and computationally expensive.
Even if linear spatial filters such as minimum mean square error (MMSE) filters are used as large-scale MUD to reduce the computational complexity, the matrix operations based on the high dimensional channel matrix are unavoidable, resulting in an increase in the circuit scale of the receiver.

To avoid such high dimensional matrix operations, fully digital two-stage beamforming has been devised in the downlink and many others \cite{Adhikary2013, Nam2014, Arvola2016, Padmanabhan2018b}.
% xi2014
%
One of the most notable methods of this approach is joint spatial division and multiplexing (JSDM)\cite{Adhikary2013}, where user equipments (UEs) are grouped based on similar transmit correlation matrices to design outer beamformer (OBF).
In practice, since the cellular UEs tend to be distributed according to the terrain around a base station (BS), UE grouping can be easily conducted based on matching angular spread on the BS side.
The OBF is created based on the long-term channel statistics, and these beams are used to effectively reduce the dimensions of equivalent channels in the angular domain\cite{Adhikary2013, Nam2014}.
Then, an inner beamformer (IBF) is applied for spatial multiplexing on the equivalent channel to manage both intra- and inter-group interference, resulting in significant reduction of the computational burden and the circuit scale of the transmitter \cite{Arvola2016, Padmanabhan2018b}.
Note that the statistics-based OBF varies over long time scales compared to the IBF, and thus requires less frequent updates.

The computational reduction and circuit scale reduction owing to the two-stage processing with the OBF can also be considered in uplink scenarios.
%
%However, the sophisticated IBF (precoder) is not available in large-scale MUD, and the spatial filtering on equivalent channel degrade detection capability
%
However, when the spatial filter is applied on the equivalent channel based on the OBF, the detection capability is significantly degraded due to severe inter-group interference (IGI) caused by the incompleteness of UE grouping (overlapping angular spread), especially when the number of UEs $M$ is the same order as the number of receive antennas $N$.
Unfortunately, the traditional massive MIMO simplification such as strong channel hardening effect \cite{Chockalingam2014} is effective and the spatial filter works well, only when $M\ll N$ is satisfied as assumed in \cite{Rusek2013, Adhikary2013}.
%This problem is greatly alleviated owing to the traditional massive MIMO simplification such as channel hardening effect \cite{Chockalingam2014} if $N>>M$ as assumed in \cite{Adhikary2013}.
%
The present study particularly focuses on high-spatial loaded MU-MIMO scenarios ($M$ in the order of $N$).
To compensate for this drawback, we introduce message passing (MP) algorithm based on belief propagation (BP) \cite{Donoho2009, Chockalingam2014, Cespedes2014, Takeuchi2017} in the post-stage of the OBF, where beliefs are exchanged between the groups as prior information in each iteration process to suppress the harmful impact of IGI.
% Rangan2016
%
%By exchanging beliefs between the groups, the suppression of harmful impacts of IGI is expected.
%By exchanging beliefs between the groups, the BP-based algorithms take advantage of the law of large numbers to converge toward near-optimal performance in large-scale MUD.

The BP-based algorithms can be roughly classified into two types: MMSE-based BP and matched filter (MF) based BP.
The most noticed MMSE-based BP is expectation propagation (EP)\cite{Cespedes2014} which is proved in \cite{Takeuchi2017} to converge toward Bayes-optimal performance under some statistical assumptions in the large system limit\cite{Chockalingam2014}. 
However, MMSE-based BP requires a predetermined number of inverse matrix operations for each transmit vector detection, which is undesirable due to prohibitively high complexity involved.
To reduce the complexity, MF-based Gaussian BP (GaBP) was presented in \cite{Chockalingam2014,Donoho2009}, but it is vulnerable to spatial fading correlations, resulting in a high-level error floor in the bit error rate (BER) performance in practical MIMO channels.

To reduce the complexity while preserving the near-optimal performance of EP \cite{Cespedes2014, Takeuchi2017}, we propose a novel design of a concatenated structure of MMSE- and MF-based BP.
%
%according to the philosophy of deep neural network (DNN).
%
The each layer is connected to each other via different activation functions, and its layered structure has some analogy to a deep neural network (DNN) \cite{John2014, Borgerding2017, Liu2017}.
%
%The potential application of deep learning (DL) based on DNN to the physical layer techniques including MUD has been increasingly recognized \cite{Wang2017, Samuel2017, Tan2018}. 
%
%In \cite{Tan2018}, the typical BP detector was re-formulated using DNN representations to estimate the system parameters using the DL. 
%
%However, the structure of BP-based DNN for MUD, which plays a major role in determining the performance, has not been studied sufficiently before learning the parameters.
%
%Therefore, we focus on how to design the DNN for MUD. 
%
As can be inferred from the structure, the proposed detector improves the convergence property of BP by efficiently suppressing the harmful impacts of intra- and inter-group interference in stages.

The remainder of this paper is organized as follows.
Sec. II presents mathematical notations and a system model as a preliminary step.
Sec. III shows the OBF design using the channel statistics to reduce the computational complexity and the circuit scale on the BS side.
The equivalent signal model in the angular domain is also defined.
A novel layered BP detector having the concatenated structure is proposed in Sec. IV.
Sec. V validates the proposed method through computer simulations and presents complexity analysis.
Finally, Sec. VI concludes the paper with a summary.

\section{System Model}
%Root ../main.tex

%===============
% Section 2.
%===============
\subsection{Mathematical Notations}

Throughout this paper, $P_{\mathsf{a}|\mathsf{b}}[a|b]$ and $p_{\mathsf{a}|\mathsf{b}}(a|b)$ respectively represent the conditional probability mass function (PMF) and the probability density function (PDF) of a realization $a$ of random variable $\mathsf{a}$ given the occurrence of a realization $b$ of random variable $\mathsf{b}$. 
$\mathbb{E}_{\mathsf{a}}\{\cdot\}$ is the expected value of random variable $\mathsf{a}$.
$\mathbb{E}_{\mathsf{a}|\mathsf{b}=b}\{\cdot\}$ denotes the conditional expectation of random variable $\mathsf{a}$ given the occurrence of a realization $b$ of random variable $\mathsf{b}$.
$\mathbb{C}^{a \times b}$ denotes a complex field of size $a \times b$.
$\mathrm{j}=\sqrt{-1}$ is the imaginary unit.
$\bm{I}_a$ represents an $a \times a$ square identity matrix.
$\mathrm{diag}\left[\bm{a}\right]$ denotes a diagonal matrix with the elements of ${\bm{a}}$.
$\mathrm{det}[\cdot]$ is determinant of a matrix.
$\underset{a\rightarrow b}{\mathrm{proj}}\left(\cdot\right)$ denotes a projection function from an $a$-dimensional vector to a $b$-dimensional vector.
$\mathrm{uniq}\{\mathcal{A}\}$ returns a set which is reconstructed with the unique elements in $\mathcal{A}$ by removing duplicate elements.

\subsection{Signal Model in the spatial domain}

In this paper, we consider uplink MUD, where the BS has $N$ receive (RX) antennas in uniform linear array (ULA) pattern and $M (\le N)$ UEs are equipped with a single TX antenna.
%
%Even though the UEs are uniformly distributed around the BS, they tend to be collocated geographically, leading to successful UE grouping from RX signal's angle of arrival (AoA) in the angular domain.
%
Here, the UEs can be grouped into $G$ group based on matching angular spread, with the set of all groups $\mathcal{G}=\left\{ 1,\ldots,g,\ldots,G \right\}$.
As $M$ and $G$ increase, it becomes inevitable that the angular spreads between adjacent groups are overlapping, leading to IGI.
Let $\mathcal{U}_g$ be the set of all UEs belonging to group $g\in\mathcal{G}$, where $|\mathcal{U}_g|=U_g$ is the number of elements in $\mathcal{U}_g$.
The set of all UEs is denoted by $\mathcal{U}=\cup_{g\in\mathcal{G}}\mathcal{U}_g$, where $|\mathcal{U}|=M=\sum_{g\in\mathcal{G}}U_g$.

The $m$-th UE conveys a modulated symbol $x_m$, which represents one of $Q$ constellation points $\mathcal{X} = \left\{\chi_1, \ldots, \chi_q, \ldots, \chi_Q \right\}$.
The average power density of the constellations in the set $\mathcal{X}$ is denoted by $E_{\mathrm{s}}$.
Let $\bm{x}=\left[x_1,\ldots,x_m,\ldots,x_M\right]^{\rm T}\in \mathcal{X}^{M\times1}$ and $\bm{y}=\left[y_1,\ldots,y_n,\ldots,y_N\right]^{\rm T}\in \mathbb{C}^{N\times1}$ denote the TX and RX symbol vectors, respectively.
Assuming frequency flat and slow fading environments, the RX vector $\bm{y}$ is given by
\begin{eqnarray}
\bm{y} = \bm{H}\bm{x}+\bm{z}.
\label{eq:y_vec}
\end{eqnarray}
$\bm{H}\in \mathbb{C}^{N\times M }$ is an $N \times M$ channel matrix in the spatial domain, where the channel fluctuations of $\bm{H}$ are static during the coherence time $T$. 
The vector $\bm{z}=[z_1,\ldots,z_n,\ldots,z_N]^{\rm T}\in \mathbb{C}^{N\times1}$ is a complex additive white Gaussian noise (AWGN) vector, whose entries $z_{n}$ obey independent identically distributed (i.i.d.) complex Gaussian distribution with zero mean and $N_0$ variance $\mathcal{CN}(0,N_0)$, where $N_{\mathrm{0}}$ is the noise power spectral density and the covariance matrix of $\bm{z}$ is given by $\mathbb{E}_{\bm{\mathsf{z}}}\left\{ \bm{z}\bm{z}^{\mathrm{H}}\right\} = N_0\bm{I}_{N}$.
With the above-mentioned signal model, the conditional PDF of RX vector $\bm{y}$ is given by
\begin{eqnarray}
p_{\bm{\mathsf{y}}|\bm{\mathsf{x}}}(\bm{y}|\bm{x})&=&\frac{1}{(\pi N_{\mathrm{0}})^{N}}\exp\left[ -\frac{\|\bm{y}-\bm{H}\bm{x}\|^2}{N_{\mathrm{0}}}\right].
\label{eq:pdf}
\end{eqnarray}

To represent the spatial fading correlation among RX antennas based on UE locations in the azimuthal direction, we use the geometric one-ring model \cite{Jakes1994}.
Assuming diffuse 2-D field of isotropic scatterers around the UEs, the element in the $i$-th row and the $j$-th column of the RX spatial correlation matrix for the $m$-th UE $\bm{\varTheta}_m\in\mathbb{C}^{N\times N}$ is given by 
\begin{eqnarray}
\left[
\bm{\varTheta}_m
\right]_{i,j}
&=&
\frac{1
}{
\bm{\Delta}\psi_m
}
\int
_{\psi_{m}^{\mathrm{min}}}
^{\psi_{m}^{\mathrm{max}}}
\exp\left[
\mathrm{j}
\pi
(i-j)
\cos\left(
\psi
\right)
\right]
d\psi,
\label{eq:correlation}
\end{eqnarray}
which denotes the correlation coefficient between the $i$-th and $j$-th RX antenna elements.
Here, waves arrive from the $m$-th UE with an angular spread $\bm{\Delta}\psi_m=\psi_m^{\mathrm{max}}-\psi_m^{\mathrm{min}}$.
The antenna element spacing is fixed to half the wavelength.
From \eqref{eq:correlation}, the RX spatial correlation becomes more severe with a decrease in the angular spread of UEs.
With the assistance of \eqref{eq:correlation}, the $m$-th column vector of $\bm{H}$ is computed by
\begin{eqnarray}
\bm{h}_m
=
\bm{\varTheta}_m^{1/2}\bm{\nu}_m,
\label{eq:channel}
\end{eqnarray}
where the each element of $\bm{\nu}_m \in\mathbb{C}^{N\times1}$ obeys $\mathcal{CN}(0,1)$.

\section{Outer Beamformer Design\\ Using Channel Statistic}
%Root ../main.tex

%===============
% Section 3.
%===============
To avoid the high dimensional matrix operations required for large-scale MUD, we adopt a two-stage receiver design consisting of the OBF and the BP-based MUD. 
The OBF determines the size of spatial filter used in the subsequent MUD, and therefore plays a major role in determining the overall computational complexity.
Let $B_g$ denotes the number of statistical beams oriented towards each UE group $g\in\mathcal{G}$, where $\sum_{g\in\mathcal{G}}B_g\le N$.
Here, the OBF for group $g$, $\bm{B}_g$, contains all statistical beams corresponding to the UEs in group $g$.
%, and the overall OBF is given by $\left[\bm{B}_1,\ldots,\bm{B}_G \right]$.
%
Assuming the UEs are grouped based on their long-term channel
statistics, e.g. channel covariance matrix, we can design group specific OBF matrices and continue to use it over multiple channel realization (over multiple coherence times).

\subsection{Outer beam selection}
There are two well known heuristic methods to find OBFs: eigen beam selection and discrete Fourier transform (DFT) beam selection \cite{Adhikary2013, Nam2014, Arvola2016, Padmanabhan2018b}.
In the eigen beam selection, the eigenvectors of the channel covariance matrix is used to form the OBF matrix via eigenvalue decomposition (EVD) \cite{Adhikary2013, Nam2014}.
By choosing $B_g$ eigenvectors corresponding to the $B_g$ largest egenvalues, we obtain the OBF matrix $\bm{B}_g$.

On the other hand, the DFT beam selection use the column vectors of DFT matrix $\bm{D}=\left[ \bm{d}_1,\ldots, \bm{d}_N \right]\in\mathbb{C}^{N\times N}$ to form the OBF matrix \cite{Arvola2016, Padmanabhan2018b}.
Then, the problem reduces to finding a subset of column vectors from the unitary DFT matrix.
In \cite{Padmanabhan2018b}, the DFT beam selection was numerically shown to be better than the eigen scheme, therefore, we use the DFT scheme to form the OBF matrix in this paper.

Let $\bm{H}_{\mathcal{U}_g}=\left[ \bm{h}_{\mathcal{U}_g(1)},\ldots,\bm{h}_{\mathcal{U}_g(U_g)} \right]$ denotes the stacked channel matrix corresponding all UEs in group $g$.
From \eqref{eq:channel}, the covariance matrix is given by
\begin{eqnarray}
\bm{R}_g
&=&
\mathbb{E}_{\mathsf{\bm{v}}}\left\{
\bm{H}_{\mathcal{U}_g}\bm{H}_{\mathcal{U}_g}^{\mathrm{H}}
\right\}
=
\sum_{i\in\mathcal{U}_g}\bm{\varTheta}_i.
\label{eq:channel_covariance}
\end{eqnarray}
To maximize the energy collected by the statistical beams, we select $B_g$ DFT column vectors that maximize the following metric for each group $g\in\mathcal{G}$ as
\begin{eqnarray}
\mathrm{Initialization:}\ 
\mathcal{D}=\{1,\ldots,N\},\ \mathcal{B}_g=\emptyset, \nonumber \\
j
=
\argmax_{\underline{n}\in\mathcal{D}}\left(
\bm{d}_{\underline{n}}^{\mathrm{H}}\bm{R}_g\bm{d}_{\underline{n}}
\right),\ \forall \underline{n}\in\mathcal{D}, \nonumber \\
\mathcal{B}_g
=
\mathcal{B}_g \cup \{ j \},\ \mathcal{D}=\mathcal{D}\setminus \mathcal{B}_g.\qquad
\label{eq:beam_selection}
\end{eqnarray}
Upon finding the subset $\mathcal{B}_g$, the resultant OBF matrix is given by $\bm{B}_g=\left[ \bm{d}_{\mathcal{B}(1)},\ldots,\bm{d}_{\mathcal{B}(B_g)} \right] \in \mathbb{C}^{N\times B_g}$ that covers the strongest signal paths of each group, where $\bm{B}_g^{\mathrm{H}}\bm{B}_g=\bm{I}_{B_g}$.

\subsection{Signal model in the angular domain}
Before shifting our focus to MUD, we represent a signal model in the angular domain by using the OBF matrix.
In each UE group $g\in\mathcal{G}$, the RX vector in the angular domain $\bm{r}_g\in\mathbb{C}^{B_g\times 1}$ can be represented as
\begin{eqnarray}
\bm{r}_g
%=
%\left[
%r_{g,1},\ldots,r_{g,b_g},\ldots,r_{g,B_g}
%\right]^{\mathrm{T}}
=
\bm{B}_g^{\mathrm{H}}\bm{y}
=
\bm{\varXi}_g\bm{x}+\bm{v}_g,
\label{eq:signal_angle1}
\end{eqnarray}
where $\bm{\varXi}_g=\bm{B}_g^{\mathrm{H}}\bm{H} \in \mathbb{C}^{B_g\times M}$ and $\bm{v}_g=\bm{B}_g\bm{z} \in \mathbb{C}^{B_g\times 1}$ are the equivalent channel and noise, respectively.
Here, $\bm{\varXi}_g$ is assumed to be perfectly estimated at the BS side in this paper\footnote{The equivalent channel $\bm{\varXi}_g$ also reduces the amount of coefficients in channel estimation, and the statistical beams improve the pilot SNR, resulting in a more accurate channel estimate. The evaluation is left for the future work.}.
However, the number of row dimensions $M$ still remains undesirable in terms of complexity.

For further dimensionality reduction of the equivalent channel matrix, we can ignore negligibly small IGI from the UEs which is located geographically remote from the group $g$, due to the attenuated sidelobes of the OBF.
In the same manner as in \eqref{eq:beam_selection}, we select $S_g\ (\le M)$ UEs that maximize the following metric for each group $g\in\mathcal{G}$ as
\begin{eqnarray}
\mathrm{Initialization:}\ 
\mathcal{M}=\{1,\ldots,M\},\ \mathcal{S}_g=\emptyset, \nonumber \\
j
=
\argmax_{\underline{m}\in\mathcal{M}}\left(
\sum_{i\in\mathcal{B}_g}\bm{d}_i^{\mathrm{H}}
\bm{\varTheta}_{\underline{m}}\bm{d}_i
\right),\ \forall \underline{m}\in\mathcal{D}, \nonumber \\
\mathcal{S}_g
=
\mathcal{S}_g \cup \{ j \},\ \mathcal{M}=\mathcal{M}\setminus \mathcal{S}_g.\qquad\qquad
\label{eq:UE_selection}
\end{eqnarray}
By ignoring IGI from the UEs not included in the subset $\mathcal{S}_g$, we can further reduce the complexity and the circuit scale.
The detailed complexity analysis is provided in Sec. V.B.

%%%%%%%%%%%%%%%
\begin{comment}
%%%%%%%%%%%%%%%
\eqref{eq:signal_angle1} can be approximately rewritten as
%
\begin{eqnarray}
\bm{r}^{(g)}
&\approx&
\bm{B}_g^{\mathrm{H}}\bm{H}_{\mathcal{S}_g}\bm{x}^{(g)}+\bm{v}^{(g)}, \nonumber \\
&=&
\grave{\bm{\varXi}}^{(g)}\grave{\bm{x}}^{(g)}+\bm{v}^{(g)},
\label{eq:signal_angle2}
\end{eqnarray}
%
where we define 
%
\begin{eqnarray}
\!\!\!\!\!\!\!\!
\grave{\bm{\varXi}}^{(g)}
\!\!\!&\!\!\triangleq\!\!&\!\!\!
%\left[ \bm{h}^{\downarrow(g)}_{1},\ldots,\bm{h}^{\downarrow(g)}_{s_g},\ldots,\bm{h}^{\downarrow(g)}_{S_g} \right]\triangleq \bm{B}_g^{\mathrm{H}}\bm{H}_{\mathcal{S}_g}, \\
\bm{B}_g^{\mathrm{H}}\bm{H}_{\mathcal{S}_g} 
=
\bm{B}_g^{\mathrm{H}}\left[
\bm{h}_{\mathcal{S}_g(1)},\ldots,\bm{h}_{\mathcal{S}_g(S_g)}
\right], \\
\!\!\!\!\!\!\!\!
\grave{\bm{x}}^{(g)}
\!\!\!&\!\!\triangleq\!\!&\!\!\!
\left[\grave{x}^{(g)}_{1},\ldots,\grave{x}^{(g)}_{s_g},\ldots,\grave{x}^{(g)}_{S_g} \right]^{\mathrm{T}}
\!\!=\!
\left[x_{\mathcal{S}_g(1)},\ldots,x_{\mathcal{S}_g(S_g)} \right]^{\mathrm{T}}\!\!\!. %\!\!\!\!\!\!
%\bm{z}^{(g)}
%\!\!\!&\!\!=\!\!&\!\!\!
%\left[z^{(g)}_{1},\ldots,z^{(g)}_{b_g},\ldots,z^{(g)}_{B_g} \right]^{\mathrm{T}}\triangleq \bm{B}_g\bm{z}.
\end{eqnarray}
%
The size of $\grave{\bm{\varXi}}^{(g)}$ reduces to $B_g\times S_g$, resulting in significant reduction of the complexity and the circuit scale.
%
The detailed complexity analysis is provided in Sec.V.
%%%%%%%%%%%%%%%
\end{comment}
%%%%%%%%%%%%%%%

\section{Design of Layered Belief Propagation}
%Root ../main.tex

%===============
% Section 4.
%===============
The dimensionality reduction of the equivalent channel based on the OBF provides vital benefits as described above, but also results in significant performance degradation in subsequent spatial filtering detection due to the incompleteness of UE grouping, as well as, due to leakage from sidelobes.
%
%Unfortunately, it becomes unavoidable as the number of concurrent UEs increases.
%
To compensate for the drawback, we introduce the BP-based iterative detection scheme.

%===========================================
% Figure 1
%===========================================
\begin{figure*}[t]
\centering
\includegraphics[width=1.8\columnwidth,keepaspectratio=true]{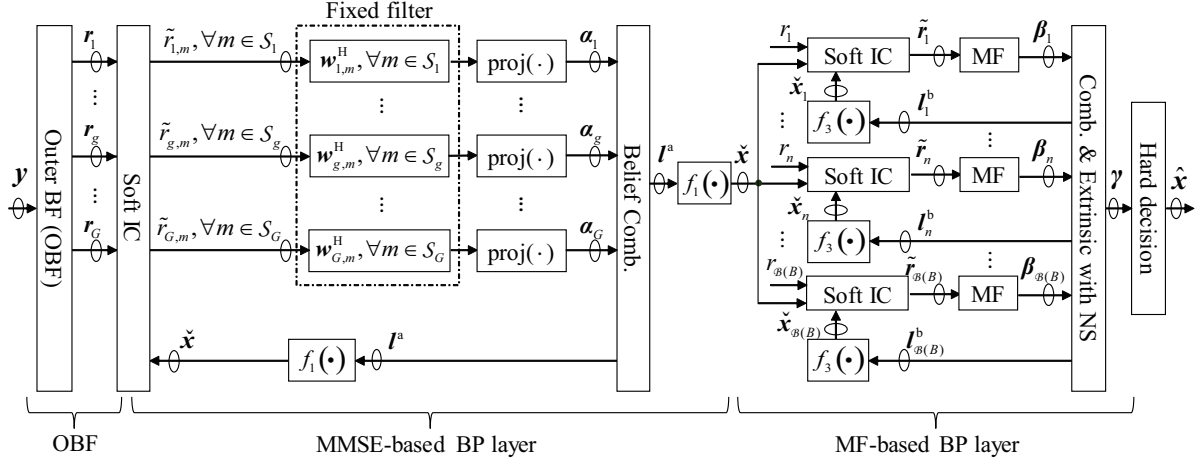}
\caption{Schematic diagram of layered BP, where the MMSE- and MF-based BP have a structure similar to DNN \cite{John2014, Borgerding2017}, respectively. Each iteration process is considered as a hidden layer and can be rewritten as linear regression model except for the activation functions in \eqref{eq:tanh} and \eqref{eq:asbtanh}. All layers are connected via the different activation functions $f_i(\cdot)\ (i=\{1,3\})$. Here, we define $\bm{\beta}_n=[\beta_{n,1},\ldots,\beta_{n,M}]^{\mathrm{T}}$, $\bm{\gamma}_n=[\gamma_{n,1},\ldots,\gamma_{n,M}]^{\mathrm{T}}$, and $\hat{\bm{x}}=[\hat{x}_{1},\ldots,\hat{x}_{M}]^{\mathrm{T}}$.
}
\label{fig:schematic}
\vspace{-2.5mm}
\end{figure*}
%===========================================

The proposed layered BP detector is illustrated in Fig. \ref{fig:schematic}.
It consists of two different layers: MMSE-based BP layer and MF-based BP layer, and the concatenated structure can efficiently suppress inra- and inter-group interference in stages.
After the MMSE-based BP layer roughly suppresses the harmful impacts of severe IGI due to the OBF, the subsequent MF-based BP layer prompts convergence of the BP algorithm by suppressing the remaining intra- and inter-group interference.
This clear role-sharing is the most vital point for reducing the complexity without sacrificing the detection capability significantly.
%
%The whole process including the preceding OBF in Fig. \ref{fig:schematic} can be regarded as a concatenated DNN with three different functions.
%
For ease of mathematical notations, we assume quadrature phase shift keying (QPSK) signaling ($\mathcal{\bm{X}}=\left\{ \pm w_x \pm \mathrm{j}w_x \right\}$, $w_x=\sqrt{E_{\mathrm{s}}/2}$)\footnote{The layered BP proposed in this paper can be extended to QAM signaling after some appropriate mathematical manipulations \cite{Takahashi2017b}.}.
%
%The group index $\cdot^{(g)}$ is omitted, and described as necessary.

\subsection{MMSE-based BP layer}

At the first iteration process ($k^{\mathrm{a}}=1$), no soft interference cancellation (Soft IC) is conducted as there are no available prior beliefs.
For each group $g \in\mathcal{G}$, upon the detection of an arbitrary TX symbol $x_m,\ \exists m\in\mathcal{S}_g$, the RX vector $\bm{r}_g$ of \eqref{eq:signal_angle1} can be approximated as
\begin{eqnarray}
\bm{r}_g
=
\tilde{\bm{r}}_{g,m}
\approx
\bm{\xi}_{g,m}x_{m}
+
\sum_{i \in \mathcal{S}_g\setminus \{m\}}
\bm{\xi}_{g,i}x_i
+
\bm{v}_g,
\label{eq:psoftcan}
\end{eqnarray}
where $\bm{\xi}_{g,m}$ denotes the $m$-th column vector of $\bm{\varXi}_g$.
In the large system limit \cite{Chockalingam2014}, the PDF of \eqref{eq:psoftcan} approaches a multivariate complex Gaussian distribution as a result of the central limit theorem; this behavior is referred to as vector Gaussian approximation (VGA).
Accordingly, $\tilde{\bm{r}}_{g,m}$ can be approximated by a Gaussian signal model as
\begin{eqnarray}
\tilde{\bm{r}}_{g,m}
=
\bm{\xi}_{g,m} x_{m} + \bm{\psi}_{g,m},
\label{eq:vga}
\end{eqnarray}
where $\bm{\psi}_{g,m}$ is the effective noise.
The covariance matrix $\bm{\varPsi}_{g,m}$ is derived from the expectation of the random variables $\bm{\mathsf{v}}_g$ and $\left\{\mathsf{x}_i,\ \forall i\in \mathcal{S}_g\setminus \{m\}\right\}$ as
\begin{eqnarray}
\bm{\varPsi}_{g,m}
&=&
\mathbb{E}_{\bm{\mathsf{v}}_g,\left\{ \mathsf{x}_i,\ \forall i\in \mathcal{S}_g\setminus \{m\} \right\}}
\left\{
\bm{\psi}_{g,m}\bm{\psi}_{g,m}^{\mathrm{H}}
\right\} \nonumber \\
%&=&
%\mathbb{E}_{\overline{\bm{\mathsf{x}}}_m | \bm{\mathsf{l}}=\bm{l}}
%\left\{
%\left[
%\grave{\bm{H}}\left(
%\overline{\bm{x}}_m-\check{\overline{\bm{x}}}_{m} \right)
%\right]
%\left[
%\grave{\bm{H}}\left(
%\overline{\bm{x}}_m-\check{\overline{\bm{x}}}_{m} \right)
%\right]^{\mathrm{H}}
%\right\} + \grave{\bm{\varPsi}} \nonumber \\
&=&
E_{\mathrm{s}}
\sum_{i\in \mathcal{S}_g \setminus \{m\}}
\bm{\xi}_{g,i}\bm{\xi}^{\mathrm{H}}_{g,i}
+N_0\bm{I}_{B_g},
\label{eq:pda_psi}
\end{eqnarray}
According to $\bm{\xi}_{g,m}$ and $\bm{\varPsi}_{g,m}$ under VGA, the PDF of $\tilde{\bm{r}}_{g,m}$ based on the Mahalanobis distance is expressed as
\begin{eqnarray}
\!\!\!\!\!&\!\!\!\!\!\!&\!\!\!\!\!\!\!
p_{\tilde{\bm{\mathsf{r}}}_{g,m} |\mathsf{x}_m}\left(
\tilde{\bm{r}}_{g,m} | x_{m}
\right)
=
\frac{1}{\pi^{B_g} \mathrm{det}\left[\bm{\varPsi}_{g,m}\right]} \nonumber \\
\!\!\!\!\!&\!\!\!\!\!\!&\!\!\!\!\!\!\!
\quad\cdot
\exp\left[ 
-\left(\tilde{\bm{r}}_{g,m}-\bm{\xi}_{g,m}x_{m}\right)^{\mathrm{H}}
\bm{\varPsi}_{g,m}^{-1}
\left(\tilde{\bm{r}}_{g,m}-\bm{\xi}_{g,m}x_{m}\right) 
\right]\!.
\label{eq:ppdf}
\end{eqnarray}
On the basis of \eqref{eq:ppdf}, a complex-valued log likelihood ratio (LLR) for Gray coded QPSK symbol $x_{m}$ is computed according to the definition \cite{Hanzo2009} as
\begin{eqnarray}
\alpha_{g,m}
\!\!&\!\!=\!\!&\!\!
\log \left[
\frac{
p_{\tilde{\bm{\mathsf{r}}}_{g,m} |\Re\left\{\mathsf{x}_{m}\right\}}
\left(
\tilde{\bm{r}}_{g,m} | \Re\left\{ x_{m} \right\} = + w_x
\right)
}{
p_{\tilde{\bm{\mathsf{r}}}_{g,m} |\Re\left\{\mathsf{x}_{m}\right\}}
\left(
\tilde{\bm{r}}_{g,m} | \Re\left\{ x_{m} \right\} = - w_x
\right)
}
\right] \nonumber \\
\!\!&\!\!\quad\!\!&\!\!
\ \ +\ 
\mathrm{j}
\log \left[
\frac{
p_{\tilde{\bm{\mathsf{r}}}_{g,m} |\Im\left\{\mathsf{x}_{m}\right\}}
\left(
\tilde{\bm{r}}_{g,m} | \Im\left\{ x_{m} \right\} = + w_x
\right)
}{
p_{\tilde{\bm{\mathsf{r}}}_{g,m} |\Im\left\{\mathsf{x}_{m}\right\}}
\left(
\tilde{\bm{r}}_{g,m} | \Im\left\{ x_{m} \right\} = - w_x
\right)
}
\right] \nonumber \\
\!\!&\!\!=\!\!&\!\!
2\sqrt{2E_{\mathrm{s}}}\ 
\bm{\xi}^{\mathrm{H}}_{g,m}\bm{\varPsi}_{g,m}^{-1}\tilde{\bm{r}}_{g,m}.
\label{eq:llr}
\end{eqnarray}
Applying the matrix inversion lemma on \eqref{eq:llr} in order to avoid computations of the matrix inversion every TX symbol detection, \eqref{eq:llr} can be rewritten as 
\begin{eqnarray}
\alpha_{g,m}
=
\bm{w}_{g,m}^{\mathrm{H}}\tilde{\bm{r}}_{g,m},
\label{eq:llr2}
\end{eqnarray}
where
\begin{eqnarray}
\bm{w}_{g,m}^{\mathrm{H}}
&=&
\frac{2\sqrt{2E_{\mathrm{s}}}}{1-E_{\mathrm{s}}\phi_{g,m}}
\bm{\xi}^{\mathrm{H}}_{g,m}
\bm{\varPsi}_{g}^{-1}, \label{eq:filter} \\
\bm{\varPsi}_{g}
&=&
E_{\mathrm{s}}
\sum_{i\in \mathcal{S}_g}
\bm{\xi}_{g,i}\bm{\xi}^{\mathrm{H}}_{g,i}
+
N_0\bm{I}_{B_g}, \label{eq:Psi}\\
\phi_{g,m}
&=&
\bm{\xi}^{\mathrm{H}}_{g,m}
\bm{\varPsi}_g^{-1}
\bm{\xi}_{g,m}.
\label{eq:parts}
\end{eqnarray}
In the same manner, the LLRs $\left\{\alpha_{g,m}, \forall m\in\mathcal{S}_g\right\}$ are computed for each group $g\in \mathcal{G}$.
To facilitate belief combining, $\alpha_{g,m}$ is projected to the $M$-dimensional vector $\bm{\alpha}_g$ as
\begin{eqnarray}
\!\!\!\!\!\!\!\!
\bm{\alpha}_g
\!\!\!&\!\!=\!\!&\!\!\!\!
\left[
\grave{\alpha}_{g,1},\ldots,\grave{\alpha}_{g,m},\ldots,\grave{\alpha}_{g,M}
\right]^{\mathrm{T}}
\!\!
=
\!\!
\underset{S_g\rightarrow M}{\mathrm{proj}}\!\!\left(\alpha_{g,m}, \forall m\in\mathcal{S}_g\right),
\end{eqnarray}
where the each element in $\bm{\alpha}_g$ is given by
\begin{eqnarray}
\grave{\alpha}_{g,m}
&=&
\begin{cases}
\alpha_{g,m} & (m\in\mathcal{S}_g) \\
0 & (\mathrm{otherwise})
\end{cases}.
\label{eq:vecllr}
\end{eqnarray}
Assuming that the effective noise included in $\bm{\alpha}_1, \ldots, \bm{\alpha}_G$ between the UE groups are not correlated, the prior belief is provided by a joint LLR vector as
\begin{eqnarray}
\bm{l}^{\mathrm{a}}
=
\left[ l_{1}^{\mathrm{a}},\ldots,l_{m}^{\mathrm{a}},\ldots,l_{M}^{\mathrm{a}} \right]^{\mathrm{T}}
=
\sum_{g\in\mathcal{G}}\bm{\alpha}_g. \label{eq:joint}
\end{eqnarray}
Finally, the soft symbol vector $\check{\bm{x}}=\left[\check{x}_{1},\ldots,\check{x}_{m},\ldots,\check{x}_{M} \right]^{\mathrm{T}}$ is obtained from the conditional expectation of $\bm{x}$ given the prior belief vector $\bm{l}^{\mathrm{a}}$ using Bayes’ rule, which is given by
\begin{eqnarray}
\check{\bm{x}}
%&=&[\check{x}_1,\dots,\check{x}_m,\ldots,\check{x}_M]^{\mathrm{T}}
%= \mathbb{E}_{\bm{\mathsf{x}} | \bm{\mathsf{l}} = \bm{l}}\left\{ \bm{x} \right\}  \nonumber \\
=\!\!
\sum_{\underline{\bm{x}}\in \mathcal{X}^{M \times 1}}
\!\!
\underline{\bm{x}}P_{\bm{\mathsf{x}} | \bm{\mathsf{l}}^{\mathrm{a}}}\left[  \underline{\bm{x}} | \bm{l}^{\mathrm{a}} \right] 
=\!\!
\sum_{\underline{\bm{x}}\in \mathcal{X}^{M \times 1}}
\!\!
\underline{\bm{x}}\frac{
p_{\bm{\mathsf{l}}^{\mathrm{a}} | \bm{\mathsf{x}}}\left( \bm{l}^{\mathrm{a}} | \underline{\bm{x}} \right)
}{
\sum_{\underline{\bm{x}}'\in \mathcal{X}^{M \times 1}}p_{\bm{\mathsf{l}}^{\mathrm{a}} | \bm{\mathsf{x}}'}\left( \bm{l}^{\mathrm{a}} | \underline{\bm{x}}' \right)
}, \label{eq:pexp}
\end{eqnarray}
where the occurrence of the vector $\bm{x}$ is assumed to be equiprobable ($P_{\bm{\mathsf{x}}}[\underline{\bm{x}}]=\frac{1}{Q^M},\ \forall \underline{\bm{x}}\in \mathcal{X}^{M\times1}$).
Moreover, assuming that the prior belief $l_m^{\mathrm{a}}$ satisfies the consistency condition\cite{Hagenauer2004}, and that its statistical behavior is uncorrelated among the elements of $\bm{l}^{\mathrm{a}}$, the entries of \eqref{eq:pexp} are readily given by \cite{Hanzo2009}
\begin{eqnarray}
\!
\check{x}_m = f_1(l_m^{\mathrm{a}})=
w_x\cdot\left(
\tanh \left[ \Re\left\{\frac{l_m^{\mathrm{a}}}{2}\right\} \right]
\!+\!
\mathrm{j}
\tanh \left[ \Im\left\{\frac{l_m^{\mathrm{a}}}{2}\right\} \right]
\right)
\label{eq:tanh},
\end{eqnarray}
where $f_1(\cdot)$ in \eqref{eq:tanh} plays the same role as an activation function (Act. func.) in DNN \cite{John2014, Borgerding2017, Liu2017}.

At the second and later iterations ($k^{\mathrm{a}}\neq 1,\ \check{\bm{x}}\neq 0$), a soft IC is carried out with the aid of the soft symbol vector $\check{\bm{x}}$ as
\begin{eqnarray}
\tilde{\bm{r}}_{g,m}
=
\bm{r}_g-\bm{\varXi}\check{\bm{x}}
+
\bm{\xi}^{\mathrm{H}}_{g,m}\check{x}_{m}.
\label{eq:psoftIC}
\end{eqnarray}
This operation can suppress the harmful impacts of severe IGI as well as the approximation errors in \eqref{eq:psoftcan}.
%
%In a typical MMSE-based BP, the spatial filter $\bm{w}_{s_g}^{\mathrm{H}}$ in \eqref{eq:filter} have to be updated according to soft IC in each iteration, resulting in high computational burden due to multiple matrix inversion operations every symbol vector transmission.

The spatial filter $\bm{w}_{g,m}^{\mathrm{H}}$ in \eqref{eq:filter} is supposed to be updated according to soft IC, leading to $\mathcal{O}(G(B_g^2S_g+B_g^3))$ complexity in each iteration for each $\bm{x}$ detection.
Although this complexity is far less than $\mathcal{O}(M^2N+M^3)$ complexity required in MMSE-based BP without the OBF, it remains high computational burden compared to the linear MMSE\footnote{The linear MMSE requires matrix operations for computing the spatial filter once in a coherence time $T$, thus, the computational complexity required for each TX vector $\bm{x}$ detection is only $O(MN)$.}.
To reduce the complexity, we may continue to use the same filter $\bm{w}_{g,m}^{\mathrm{H}}$ in all iterations. 
In this case, the dominant factor in determining the complexity is soft IC in \eqref{eq:psoftIC}, which has $\mathcal{O}(MN)$ complexity (in the same order of the linear MMSE).
Of course, the fixed $\bm{w}_{g,m}^{\mathrm{H}}$ is not matched to the residual interference in the iteration process, therefore, this approximation induces the model error of beliefs and deteriorates the convergence property of BP.
%
%However, the harmful impacts of IGI caused by the dimensional reduction based on the OBF can be sufficiently suppressed owing to iterative filtering process with the aid of soft IC.

After updating the LLR $\alpha_{g,m}$ in \eqref{eq:llr2} with $\tilde{\bm{r}}_{g,m}$ in \eqref{eq:psoftIC}, equations \eqref{eq:vecllr}, \eqref{eq:joint} and \eqref{eq:tanh} are computed again in each iteration.
When the number of iterations reaches the predetermined value $K^{\mathrm{a}}$, the symbol symbol $\check{\bm{x}}$ is used as input to the subsequent MF-based BP as the prior information.
%
%\begin{eqnarray}
%\check{\bm{X}}
%=
%\mathrm{repmat}\left( \check{\bm{x}}, N \right).
%\label{eq:dec}
%\end{eqnarray}

\subsection{MF-based BP layer}

To prompt convergence of the BP algorithm while mitigating a model error of beliefs due to the fixed filter approximation in the preceding process, Gaussian BP (GaBP) \cite{Chockalingam2014, Donoho2009} is used as MF-based BP.

For further computational reduction, a simple MF belief (MF-GaBP) \cite{takahashi2018} is utilized, instead of a traditional LLR belief (LLR-GaBP) \cite{Chockalingam2014}.
The all processes in MF-GaBP except for belief combining is individually conducted for each DFT beam selected in \eqref{eq:beam_selection}, therefore, duplicate elements in $\cup_{g\in\mathcal{G}}\mathcal{B}_g$ should be removed in advance.
Thus, we define the set $\mathcal{B}=\mathrm{uniq}\left\{\cup_{g\in\mathcal{G}}\mathcal{B}_g\right\}$, where $|\mathcal{B}|=B$.
%
%%%%%%%%%%%%%%%
\begin{comment}
%%%%%%%%%%%%%%%
However, when MF-GaBP is utilized for the receive signal in \eqref{eq:signal_angle1}, the belief combining is not successfully processed when $\bigcup_{g\in\mathcal{G}}\mathcal{B}_g$ have duplicate elements.
%
To avoid the impairment, we define the set $\mathcal{B}=\mathrm{uniq}\left[ \bigcup_{g\in\mathcal{G}}\mathcal{B}_g \right]$, and the following receive signal:
%
\begin{equation}
\bm{\lambda}
=
\mathrm{uniq}
\left[
\bm{r}^{(1)\mathrm{T}},\ldots,\bm{r}^{(G)\mathrm{T}}
\right]^{\mathrm{T}}
=
\acute{\bm{\varXi}}\bm{x}+\acute{\bm{v}},
\label{eq:signal_model3}
\end{equation}
%
where $\acute{\bm{\varXi}}\in\mathcal{C}^{B\times M}$ is given by vertically stacking the row vector of $\left[ \bm{\varXi}^{(1)},\ldots,\bm{\varXi}^{(G)}\right]$ in \eqref{eq:signal_angle1} corresponding to the elements in $\bm{\lambda}$, and $\acute{\bm{v}}$ is similarly given.
%
$B$ denotes the number of elements in $\bm{\lambda}$.
%
Note that MF-GaBP consists only of scalar operations, thus, the size of spatial filter is no longer vital.
%
Therefore, the algorithm is derived based on \eqref{eq:signal_model3}.
%%%%%%%%%%%%%%%
\end{comment}
%%%%%%%%%%%%%%%

\subsubsection{BP with Activation Function}

First, the soft IC for the $n$-th RX symbol $r_n=\bm{d}_n^{\mathrm{H}}\bm{y}$, is carried out with the aid of a soft symbol vector $\check{\bm{x}}_{n} = \left[\check{x}_{n,1}, \ldots, \check{x}_{n,m}, \ldots, \check{x}_{n,M} \right]^{\rm T}$.
At the first iteration process ($k^{\mathrm{b}}=1$), the soft symbol vector is given by $\check{\bm{x}}_n=\check{\bm{x}},\ \forall n\in\mathcal{B}$.
In the detection of an arbitrary TX symbol $x_m,\ \exists m\in\{1,\ldots,M\}$, the cancellation process is expressed as 
\begin{equation}
\tilde{r}_{n,m}
=
r_{n} - \bm{\xi}_{n}^{\rightarrow} \check{\bm{x}}_n + \xi_{n,m} \check{x}_{n,m},\ \forall n\in\mathcal{B},
\label{eq:gsoftIC}
\end{equation}
where $\bm{\xi}_{n}^{\rightarrow}=\bm{d}_n^{\mathrm{H}}\bm{H}=[\xi_{n,1},\ldots,\xi_{n,m},\ldots,\xi_{n,M}]$ is the row vector of the equivalent channel.
%
%For each elements in $\bm{\lambda}$, the output vector of soft IC $\tilde{\bm{\lambda}}_b= [ \tilde{\lambda}_{b,1}, \ldots,\tilde{\lambda}_{b,m},\ldots,\tilde{\lambda}_{b,M}]^\mathrm{T}$ is computed.

In MF-GaBP, instead of traditional LLR, we utilize MF output as a belief, which is given by \cite{takahashi2018}
\begin{eqnarray}
\beta_{n,m}
&=&
\xi_{n,m}^*\tilde{r}_{n,m},\ \forall n\in\mathcal{B}. \label{eq:beta}
\end{eqnarray}
Assuming that effective noise included in $\beta_{1,m},\ldots,\beta_{\mathcal{B}(B),m}$ are not correlated, a posterior belief is provided by a maximum ratio combining (MRC) of $\beta_{n,m}$ as
\begin{eqnarray}
\gamma_{m}&=&\sum_{i\in\mathcal{B}}\beta_{i,m}. \label{eq:post}
\end{eqnarray}
Nevertheless, if $\gamma_{m}$ is utilized as the prior belief, GaBP is subject to ill convergence behavior of iterative detection, due to the correlation between $r_n$ and $\beta_{n,m}$ included in $\gamma_m$.
In GaBP regime \cite{Chockalingam2014, Donoho2009}, therefore, the prior belief is typically provided by an extrinsic belief $l_{n,m}^{\mathrm{b}}$, which is given by
\begin{eqnarray}
\l_{n,m}^{\mathrm{b}} = \gamma_m-\beta_{n,m}=\sum_{i\in\mathcal{B}\setminus \{n\}}\beta_{i,m}. \label{eq:mfbelief}
\end{eqnarray}
Here, a prior belief vector $\bm{l}_n^{\mathrm{b}}=\left[ l_{n,1}^{\mathrm{b}}, \ldots, l_{n,m}^{\mathrm{b}}, \ldots, l_{n,M}^{\mathrm{b}} \right]^{\mathrm{T}}$ is constructed by using \eqref{eq:mfbelief}.

Finally, the soft symbol vector $\check{\bm{x}}_n$ is obtained from the conditional expectation of $\bm{x}$ given the prior belief vector $\bm{l}_n^{\mathrm{b}}$ as in the same manner of \eqref{eq:pexp}.
%
%\begin{eqnarray}
%\check{\bm{x}}_b
%=\!\!
%\sum_{\underline{\bm{x}}\in \mathcal{X}^{M \times 1}}
%\!\!
%\underline{\bm{x}}\frac{
%p_{\bm{\mathsf{l}}_b^{\mathrm{b}} | \bm{\mathsf{x}}}\left( \bm{l}_b^{\mathrm{b}} | \underline{\bm{x}} %\right)
%}{
%\sum_{\underline{\bm{x}}'\in \mathcal{X}^{M \times 1}}p_{\bm{\mathsf{l}}_b^{\mathrm{b}} | %\bm{\mathsf{x}}}\left( \bm{l}_b^{\mathrm{b}} | \underline{\bm{x}}' \right)
%}. \label{eq:gexp}
%\end{eqnarray}
%
However, $l_{n,m}^{\mathrm{b}}$ is no longer the LLR belief but the MF belief, thus, we are not able to use \eqref{eq:tanh} to compute the conditional expectation.
To avoid the impairment, we should focus on strong channel hardening effect \cite{Chockalingam2014} in the large system limit.
Then, the soft symbol can be approximately computed by the following Act. func.: \cite{takahashi2018}
\begin{eqnarray}
\!\!\!\!\!\!&\!\!\!\!\!\!&\!\!\!\!\!\!\!\!\!\!\!\!
\check{x}_{n,m}
=
f_2(l_{n,m}^{\mathrm{b}}) \nonumber \\
    \!\!\!\!\!\!&\!\!\!\!\!\!&\!\!\!
=
w_x\cdot\left(
\tanh\! \left[\! \frac{\sqrt{2E_s}}{N_0}\Re\!\left\{l_{n,m}^{\mathrm{b}}\right\} \!\right]
\!+\!
\mathrm{j}
\tanh\! \left[\! \frac{\sqrt{2E_s}}{N_0}\Im\!\left\{l_{n,m}^{\mathrm{b}}\right\} \!\right]
\right)\!,
\label{eq:mftanh}
\end{eqnarray}
instead of \eqref{eq:tanh}.
Unfortunately, it is difficult to satisfy the large system limit in practical large-scale MUD scenarios, due to the limitation of the receiver and the spatial fading correlation.
This approximation error results in severe error propagation, leading to ill convergence behavior of iterative detection. 
%
%This approximation error induces severe error propagation, leading to abnormal noise enhancements in the prior beliefs in the iterative process.
%
To deal with the inconvenience, adaptive scaled belief (ASB) was proposed in \cite{takahashi2018}, and this technique is equivalent to using the following Act. func. to generate soft symbols:
\begin{eqnarray}
\!\!\!\!\!\!&\!\!\!\!\!\!&\!\!\!\!\!\!\!\!\!\!\!\!
\check{x}_{n,m}
=
f_3(l_{n,m}^{\mathrm{b}}) \nonumber \\
    \!\!\!\!\!\!&\!\!\!\!\!\!&\!\!\!
=
w_x\cdot\left(
\tanh\! \left[\! \frac{\sigma}{\omega_{n,m}}\Re\!\left\{l_{n,m}^{\mathrm{b}}\right\} \!\right]
\!+\!
\mathrm{j}
\tanh\! \left[\! \frac{\sigma}{\omega_{n,m}}\Im\!\left\{l_{n,m}^{\mathrm{b}}\right\} \!\right]
\right)\!,
\label{eq:asbtanh}
\end{eqnarray}
instead of \eqref{eq:mftanh}, where
\begin{eqnarray}
\omega_{n,m}
&=&
w_x\sum_{i\in\mathcal{B}\setminus \{n\}}|\xi_{i,m}|^2
\label{eq:omega}.
\end{eqnarray}
$\sigma$ is the predetermined design parameter; in the simulations, we set $\sigma=2.0$.
In this paper, we utilize $f_3(\cdot)$ in \eqref{eq:asbtanh} as Act. func..
When the number of iterations in MF-GaBP reaches the predetermined value $K^{\mathrm{b}}$, $x_m$ is detected by
\begin{eqnarray}
\hat{x}_{m}&=&
w_x\cdot \left(
\mathrm{sgn}\left[
\Re \left\{ \gamma_{m} \right\}
\right]
+
\mathrm{j}\ \mathrm{sgn}\left[
\Im \left\{ \gamma_{m} \right\}
\right]
\right), \label{eq:dec}
\end{eqnarray}
where $\mathrm{sgn}[\cdot]$ denotes the operation for extracting the sign of a number.

\subsubsection{Node Selection Method}

In the presence of spatial fading correlation on the BS side, $\beta_{1,m},\ldots,\beta_{\mathcal{B}(B),m}$ are highly correlated, and the belief combining is not successfully processed in \eqref{eq:mfbelief}.
To suppress the harmful effects of the model error, we use the node selection (NS) method \cite{takahashi2018}, in which $\beta_{1,m},\ldots,\beta_{\mathcal{B}(B),m}$ are divided into several sets so that the beliefs in the same set are less-correlated as much as possible based on the RX antenna pattern, and sequentially updated for each sets.
The NS method is discussed in detail in \cite{takahashi2018}.
The number of sets is $4$ in this paper.
Note that the NS method does not change the complexity of MF-GaBP.
%In the proposed method, two different BP with different roles are concatenated, in which the each layer is connected to each other via different activation functions.
%
%Including the fact that the each layer has different roles, its structure is extremely similar to DNN.

\section{Numerical Results}
%Root ../main.tex

%===============
% Section 6.
%===============
\subsection{BER performance}

Computer simulations were conducted to validate the performance of the proposed layered BP detector for large-scale MUD based on the OBF, where the performance metrics are averaged over 1000 independent UE drops and channel realizations.
The average RX power from each TX antenna is assumed to be identical on the basis of slow TX power control.
A sector antenna of $120$ degrees opening is considered.
The UEs are naturally partitioned into $G$ segments with $M/G$ UEs randomly dropped in each segment.
The harmful impacts of IGI become more severe as $M$ and $G$ increase.
The angular spread for each UE is $15$ degrees.
The modulation scheme is Gray-coded QPSK.
Time and frequency synchronization between TX and RX are assumed to be perfect.

%===========================================
% Figure 2 (BER)
%===========================================
\begin{figure}[!t]
\centering
\subfigure[$(N,M)=(64,40)$, $U_g=M/G=5,\ \forall g\in\mathcal{G}$]{
\includegraphics[width=8cm]{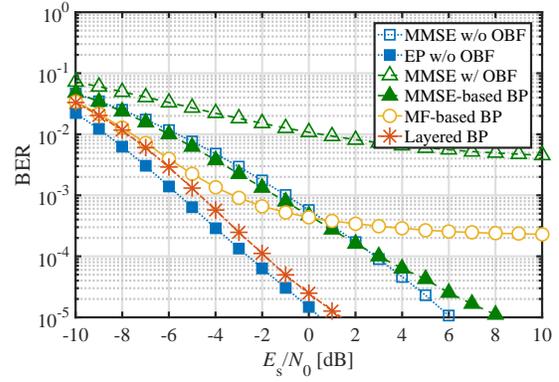}
\label{fig:ber40}}
\subfigure[$(N,M)=(64,48)$, $U_g=M/G=6,\ \forall g\in\mathcal{G}$]{
\includegraphics[width=8cm]{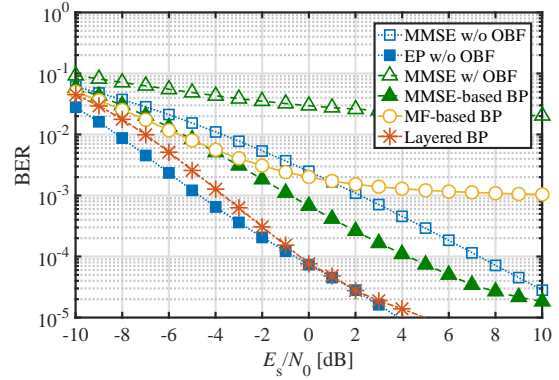}
\label{fig:ber48}}
\caption{BER performances with $(G,B_g,S_g)=(8,8,16),\ \forall g\in\mathcal{G}$.}
\label{fig:ber}
\vspace{-5mm}
\end{figure}
%===========================================

Fig. 2 shows BER performances with the system parameters $(G,B_g,S_g)=(8,8,16),\ \forall g\in\mathcal{G}$ in $(N,M)=(64,40)$ and $(N,M)=(64,48)$ MU-MIMO configurations, respectively.
The following performances are compared:
\begin{itemize}
    \item MMSE w/o OBF\ :\ is drawn as a baseline performance.
    \item EP w/o OBF\ :\ is drawn as a comparison with the leading-edge BP detector \cite{Cespedes2014, Takeuchi2017}. The damping factor is $0.2$ as shown in \cite{Cespedes2014}, where the number of iterations is $K=16$. The filtering process is conducted across all antennas.
    \item MMSE w/ OBF\ :\ is the MMSE detector with the OBF.
    \item MMSE-based BP\ :\ is MMSE-based BP with fixed filter $\bm{w}_g^{\mathrm{H}}$ presented in Sec. IV.A, which is given by stopping the layered BP before MF-based BP, where $K^{\mathrm{a}}=16$. 
    %\item MMSE-based BP (update)\ :\ is MMSE-based BP when $\bm{w}_g^{\mathrm{H}}$ is updated in \eqref{eq:filter} every iteration, where $K^{\mathrm{a}}=8$. 
    \item MF-based BP\ :\ is MF-GaBP presented in Sec. IV.B, which is given by setting $\check{\bm{x}}_n=\bm{0},\ \forall n\in\mathcal{B}$ in \eqref{eq:gsoftIC} at the first iteration, where $K^{\mathrm{b}}=16$.
    \item Layered BP: is the proposed concatenated detector presented in Sec. IV, where $(K^{\mathrm{a}},K^{\mathrm{b}})=(8,8)$.  
\end{itemize}
Compared to ``MMSE w/o OBF'', the performances of ``MMSE w/ OBF'' are significantly degraded due to the dimensionality reduction based on the OBF.
To compensate for the drawback, we introduced the BP detectors.
Owing to the iterative filtering process using the fixed filter with the aid of soft IC, ``MMSE-based BP'' can suppress the harmful impacts of severe IGI and achieve BER = $10^{-4}$ in both cases.
Unfortunately, the fixed filter cannot capture the change in the statistical model of beliefs in iterative process, therefore, there are still large performance gaps between ``MMSE-based BP'' and ``EP w/o OBF''.
On the other hand, ``MF-based BP'' without matrix operations alone cannot sufficiently suppress the impairments for the OBF and the high-level error floor is observed.
%
%On the other hand, when we accept $\mathcal{O}(G(B_g^2S_g+B_g^3))$ complexity to update $\bm{w}_g^{\mathrm{H}}$ in \eqref{eq:filter} every iteration, ``MMSE-based BP (update)'' can approach the ``EP w/o OBF'' performance.
%
%An appropriate design of BP is able to compensate for the drawback due to the dimensional reduction based on OBF.

By concatenating these two BP detectors with different functions, these inconveniences can be improved in stages.
The most attractive feature is that our proposed ``Layered BP'' can approach the ``EP w/o OBF'' performance without matrix operations in each iteration.
Remarkably, the degradation is less than 1.0 dB in $M=40$ and lass than 0.5 dB in $M=48$ at BER = $10^{-4}$.
The total gain from ``MMSE w/o OBF'' is about $5.0$ dB and $7.0$ dB at BER = $10^{-4}$, respectively.
%
%In fact, it is difficult to obtain the above performances when  we change the configuration and order of filters, and activation functions.
%
%These results imply that we can dramatically reduce the computational complexity without sacrificing much on the performance by suppressing errors in stages owing to the concatenated structure.

\subsection{Complexity analysis}

First, we focus on the operations required once per channel realization (per coherence time $T$).
Note that we can ignore the complexity for the OBF design as it can be computed off-line.
In the cases of ``w/o OBF'', it is necessary to compute matrix inversion and matrix multiplications according to the channel size $N\times M$, leading to $\mathcal{O}(M^2N+M^3)$ complexity.
On the other hand, the OBF can reduce the equivalent channel size to $B_g\times S_g$, thus, the complexity for computing $\bm{w}_g^{\mathrm{H}},\ \forall g\in\mathcal{G}$ in \eqref{eq:filter} is $\mathcal{O}(G(B_g^2S_g+B_g^3))$.
As an example, when $(G,B_g,S_g)=(8,8,16),\ \forall g\in\mathcal{G}$ in $(N,M)=(64,48)$, its complexity is roughly $(G(B_g^2S_g+B_g^3))/(M^2N+M^3)=1/21\simeq 5\%$ fraction of the cases of ``w/o OBF''.
As the most attractive point, the reduction of filter size allows us to significantly reduce the circuit scale of the receiver.

%%%% Table 1 %%%
\begin{table}[t]
\caption{Computational Comparison of different detectors}
\label{tab:comp}
\begin{center}
\vspace{-2.5mm}
\scalebox{0.9}{
\begingroup
\renewcommand{\arraystretch}{1.4}
\begin{tabular}{|c|c|}\hline
Detector & The order of complexity for each $\bm{x}$ detection \\ \hline
MMSE w/o OBF    & $\mathcal{O}(MN)$\\ \hline
EP w/o OBF      & $\mathcal{O}(K(M^2N+M^3))$\\ \hline
MMSE w/ OBF     & $\mathcal{O}(GB_gS_g)$\\ \hline 
MF-based BP     & $\mathcal{O}(K^{\mathrm{b}}MN)$\\ \hline
MMSE-based BP   & $\mathcal{O}(K^{\mathrm{a}}MN)$\\ \hline
%MMSE-based BP (update)  & $\mathcal{O}(K( G(B_g^2S_g + B_g^3) + MN))$\\ \hline
Layered BP      & $\mathcal{O}((K^{\mathrm{a}}+K^{\mathrm{b}})\ MN)$\\ 
\hline
\end{tabular}
\endgroup
}
\vspace{-6mm}
\end{center}
\end{table}

%===========================================
% Figure 3
%===========================================
\begin{figure}[t]
\centering
\includegraphics[width=0.85\columnwidth,keepaspectratio=true]{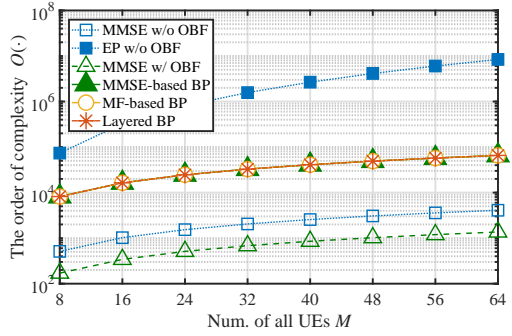}
\caption{The order of complexities as a function of number of all UEs $M$ (from 8 to 64), whereas that of RX antennas $N$ is fixed at 64. The system parameters $(G,B_g,S_g)=(8,N/G,\lceil M/3 \rceil),\ \forall g\in\mathcal{G}$. The number of iterations is $K=16$ in ``EP w/o OBF'', $K^{\mathrm{a}}=16$ in ``MMSE-based BP'', $K^{\mathrm{b}}=16$ in ``MF-based BP'', and $(K^{\mathrm{a}},K^{\mathrm{b}})=(8,8)$ in ``Layered BP'', respectively.}
\vspace{-5mm}
\label{fig:complexity}
\end{figure}
%===========================================

Let shift our focus to the computational complexity required for each TX vector $\bm{x}$ detection in order to evaluate the complexity of iterative detection.
Table \ref{tab:comp} summarizes the order of complexities, and they are plotted under some system conditions in Fig. \ref{fig:complexity}.
Due to the matrix operations according to $N\times M$ channel matrix in each iteration, the complexity of ``EP w/o OBF'' rapidly grows as $M$ increases.
Owing to the dimensionality reduction based on the OBF, by contrast, the other BP detectors can significantly reduce the complexity.
In particular, the complexity of ``Layered BP'', which mainly involves the soft IC in \eqref{eq:psoftIC} and \eqref{eq:gsoftIC} and filtering process in \eqref{eq:llr2} and \eqref{eq:beta}, is $\mathcal{O}(MN)$ in each iteration, which is sufficiently acceptable complexity if the number of iterations $K^{\mathrm{a}}+K^{\mathrm{b}}$ is about 10 to 20.
Surprisingly, from Figs. \ref{fig:ber} and \ref{fig:complexity}, we can found that ``Layered BP'' can achieve almost the same performance as ``EP w/o OBF'' with less than $2\%$ complexity of ``EP w/o OBF''.
These results imply that we can dramatically reduce the computational complexity without sacrificing much on the performance by suppressing errors in stages owing to the concatenated layered structure.

\section{Conclusion}
%Root ../main.tex

%===============
% Section 7.
%===============
\vspace{-1mm}
In this paper, we proposed a novel layered BP detector having the concatenated structure of two different BP layers for low-complexity large MUD based on the OBF. 
To reduce the computational complexity for MUD, the OBF designed based on the long term channel statistics reduces the number of dimensions of the equivalent channel, leading to significant reduction of the circuit scale on the BS side.
As $M$ and $G$ increases, however, a high-precision detection in the subsequent MUD becomes difficult due to the incompleteness of UE grouping, as well as, due to leakage from sidelobes.
The concatenated structure of the layered BP is able to improve the detection capability by suppressing the model error of beliefs caused by intra- and inter-group interference, in stages.
Through computer simulations, we can confirm that the layered BP detector dramatically reduce the complexity without sacrificing the detection capability significantly.
%
%The design of the BP detector based on the philosophy of DNN might open new vistas for deployments of large-scale MUD.

%\vspace{-0.25mm}
\section*{Acknowledgement}
%\vspace{-0.5mm}
This work was financially supported by JSPS KAKENHI Grant Number JP18H03765, Japan and the Osaka University Scholarship for Overseas Research Activities 2018.
%\vspace{-0.25mm}

\bibliographystyle{IEEEtran}
\bibliography{IEEEabrv,conf_abbrv,ref}

\begin{thebibliography}{10}
\providecommand{\url}[1]{#1}
\csname url@rmstyle\endcsname
\providecommand{\newblock}{\relax}
\providecommand{\bibinfo}[2]{#2}
\providecommand\BIBentrySTDinterwordspacing{\spaceskip=0pt\relax}
\providecommand\BIBentryALTinterwordstretchfactor{4}
\providecommand\BIBentryALTinterwordspacing{\spaceskip=\fontdimen2\font plus
\BIBentryALTinterwordstretchfactor\fontdimen3\font minus
  \fontdimen4\font\relax}
\providecommand\BIBforeignlanguage[2]{{%
\expandafter\ifx\csname l@#1\endcsname\relax
\typeout{** WARNING: IEEEtran.bst: No hyphenation pattern has been}%
\typeout{** loaded for the language `#1'. Using the pattern for}%
\typeout{** the default language instead.}%
\else
\language=\csname l@#1\endcsname
\fi
#2}}

\bibitem{Hanzo2009}
L.~Hanzo, O.~Alamri, M.~El-Hajjar, and N.~Wu, \emph{Near-Capacity
  Multi-Functional {MIMO} Systems Sphere-Packing, Iterative Detection and
  Cooperation}.\hskip 1em plus 0.5em minus 0.4em\relax Wiley-IEEE Press, May
  2009.

\bibitem{yang2015}
S.~Yang and L.~Hanzo, ``Fifty years of {MIMO} detection: The road to
  large-scale {MIMO}s,'' \emph{IEEE Commun. Surveys Tutorials}, vol.~17, no.~4,
  pp. 1941--1988, Fourthquarter 2015.

\bibitem{Rusek2013}
F.~Rusek, D.~Persson, B.~K. Lau, E.~G. Larsson, T.~L. Marzetta, O.~Edfors, and
  F.~Tufvesson, ``Scaling up {MIMO}: Opportunities and challenges with very
  large arrays,'' \emph{IEEE Signal Processing Magazine}, vol.~30, no.~1, pp.
  40--60, Jan. 2013.

\bibitem{Adhikary2013}
A.~Adhikary, J.~Nam, J.~Ahn, and G.~Caire, ``Joint spatial division and
  multiplexing—{T}he large-scale array regime,'' \emph{IEEE Trans. Inf.
  Theory}, vol.~59, no.~10, pp. 6441--6463, Oct 2013.

\bibitem{Nam2014}
J.~Nam, A.~Adhikary, J.~Ahn, and G.~Caire, ``Joint spatial division and
  multiplexing: Opportunistic beamforming, user grouping and simplified
  downlink scheduling,'' \emph{IEEE Journal of Selected Topics in Signal
  Processing}, vol.~8, no.~5, pp. 876--890, Oct. 2014.

\bibitem{Arvola2016}
A.~Arvola, A.~T{\"o}lli, and D.~Gesbert, ``Two-layer precoding for
  dimensionality reduction in massive {MIMO},'' in \emph{2016 24th European
  Signal Processing Conf. (EUSIPCO)}, Aug 2016, pp. 2000--2004.

\bibitem{Padmanabhan2018b}
A.~Padmanabhan and A.~T{\"o}lli, ``An iterative approach for inter-group
  interference management in two-stage precoder design,'' in \emph{Proc. IEEE
  Globecom 2018, Abu Dhabi, UAE}, Dec. 2018.

\bibitem{Chockalingam2014}
A.~Chockalingam and B.~S. Rajan, \emph{Large {MIMO} {S}ystems}.\hskip 1em plus
  0.5em minus 0.4em\relax Cambridge University Press, 2014.

\bibitem{Donoho2009}
D.~L. Donoho, A.~Maleki, and A.~Montanari, ``Message-passing algorithms for
  compressed sensing,'' \emph{Proc. Nat. Acad. Sci.}, vol. 106, no.~45, pp.
  18\,914--18\,919, Nov. 2009.

\bibitem{Cespedes2014}
J.~\'Cespedes, P.~M. Olmos, M.~S\'anchez-Fern\'andez, and F.~Perez-Cruz,
  ``Expectation propagation detection for high-order high-dimensional {MIMO}
  systems,'' \emph{IEEE Trans. Commun.}, vol.~62, no.~8, pp. 2840--2849, Aug.
  2014.

\bibitem{Takeuchi2017}
\BIBentryALTinterwordspacing
K.~Takeuchi, ``Rigorous dynamics of expectation-propagation-based signal
  recovery from unitarily invariant measurements,'' Jan. 2017. [Online].
  Available: \url{http://arxiv.org/abs/1701.05284}
\BIBentrySTDinterwordspacing

\bibitem{John2014}
\BIBentryALTinterwordspacing
J.~R. Hershey, J.~L. Roux, and F.~Weninger, ``Deep unfolding: Model-based
  inspiration of novel deep architectures,'' \emph{CoRR}, vol. abs/1409.2574,
  2014. [Online]. Available: \url{http://arxiv.org/abs/1409.2574}
\BIBentrySTDinterwordspacing

\bibitem{Borgerding2017}
M.~Borgerding, P.~Schniter, and S.~Rangan, ``{AMP}-inspired deep networks for
  sparse linear inverse problems,'' \emph{IEEE Trans. Signal Processing},
  vol.~65, no.~16, pp. 4293--4308, Aug. 2017.

\bibitem{Liu2017}
W.~Liu, Z.~Wang, X.~Liu, N.~Zeng, Y.~Liu, and F.~E. Alsaadi, ``A survey of deep
  neural network architectures and their applications,'' \emph{Neurocomputing},
  vol. 234, pp. 11--26, 2017.

\bibitem{Jakes1994}
W.~C. Jakes, \emph{Microwave Mobile Communications, 1st ed.}\hskip 1em plus
  0.5em minus 0.4em\relax Wiley-IEEE Press, 1994.

\bibitem{Takahashi2017b}
T.~Takahashi, S.~Ibi, and S.~Sampei, ``Design of adaptively scaled belief in
  large {MIMO} detection for higher-order modulation,'' in \emph{2017 APSIPA
  ASC}, Dec. 2017, pp. 1800--1505.

\bibitem{Hagenauer2004}
J.~Hagenauer, ``The {EXIT} chart - introduction to extrinsic information
  transfer,'' in \emph{Proc. EUSIPCO 2004}, Vienna, Austria, Sept. 2004, pp.
  1541--1548.

\bibitem{takahashi2018}
T.~Takahashi, S.~Ibi, and S.~Sampei, ``Design of criterion for adaptively
  scaled belief in iterative large {MIMO} detection,'' \emph{IEICE Trans.
  Commun.}, vol. E102-B, no.~2, 2018.

\end{thebibliography}
%\input{main.bbl}

\end{document}